\documentclass[twocolumn]{aastex62} 
\usepackage[utf8]{inputenc}
\usepackage[T1]{fontenc}
\usepackage{wrapfig}
\usepackage{listings}
\usepackage{amsmath}
\usepackage{amssymb}
\usepackage{multirow}
\usepackage{hyperref}
\usepackage{natbib}

\definecolor{mygray}{rgb}{0.95,0.95,0.95}

\newcommand\axs{{\em AXS}}

\shorttitle{Astronomy eXtensions for Spark}
\shortauthors{Ze\v{c}evi\'c et al.}

\begin{document}
\title{AXS: A framework for fast astronomical data processing based on Apache Spark}

\correspondingauthor{Petar Ze\v{c}evi\'c}
\email{petar.zecevic@fer.hr}

\author[0000-0002-2651-243X]{Petar Ze\v{c}evi\'c}
\affiliation{Faculty of Electrical Engineering and Computing, University of Zagreb, Croatia}
\affiliation{Visiting Fellow, DIRAC Institute, University of Washington, Seattle, USA}

\author[0000-0002-0558-0521]{Colin T. Slater}
\affiliation{DIRAC Institute and the Department of Astronomy, University of Washington, Seattle, USA}

\author[0000-0003-1996-9252]{Mario Juri\'c}
\affiliation{DIRAC Institute and the Department of Astronomy, University of Washington, Seattle, USA}

\author[0000-0001-5576-8189]{Andrew J. Connolly}
\affiliation{DIRAC Institute and the Department of Astronomy, University of Washington, Seattle, USA}

\author[0000-0002-4857-5351]{Sven Lon\v{c}ari\'c}
\affiliation{Faculty of Electrical Engineering and Computing, University of Zagreb, Croatia}

\author[0000-0003-1996-9253]{Eric C. Bellm}
\affiliation{DIRAC Institute and the Department of Astronomy, University of Washington, Seattle, USA}

\author[0000-0003-1996-9254]{V. Zack Golkhou}
\affiliation{DIRAC Institute and the Department of Astronomy, University of Washington, Seattle, USA}

\author[0000-0003-1996-9255]{Krzysztof Suberlak}
\affiliation{DIRAC Institute and the Department of Astronomy, University of Washington, Seattle, USA}

\begin{abstract}

We introduce AXS (Astronomy eXtensions for Spark), a scalable open-source astronomical data analysis framework built on Apache Spark, a widely used industry-standard engine for big data processing. Building on capabilities present in Spark, AXS aims to enable querying and analyzing almost arbitrarily large astronomical catalogs using familiar Python/AstroPy concepts, DataFrame APIs, and SQL statements. We achieve this by i) adding support to Spark for efficient on-line positional cross-matching and ii) supplying a Python library supporting commonly-used operations for astronomical data analysis. To support scalable cross-matching, we developed a variant of the ZONES algorithm \citep{there-goes_gray_2004} capable of operating in distributed, shared-nothing architecture. We couple this to a data partitioning scheme that enables fast catalog cross-matching and handles the data skew often present in deep all-sky data sets. The cross-match and other often-used functionalities are exposed to the end users through an easy-to-use Python API. We demonstrate AXS' technical and scientific performance on SDSS, ZTF, Gaia DR2, and AllWise catalogs. Using AXS we were able to perform on-the-fly cross-match of Gaia DR2 (1.8 billion rows) and AllWise (900 million rows) data sets in $\sim 30$ seconds. We discuss how cloud-ready distributed systems like AXS provide a natural way to enable comprehensive end-user analyses of large datasets such as LSST.
\end{abstract}


\section{Introduction}
\label{sec:org197424c}
The amount of astronomical data available for analysis is growing at an ever increasing rate. For example, the 2MASS survey (in operation 1997-2003) delivered a catalog containing 470 million sources (about 40 GB). More recently, the latest release (DR14) of Sloan Digital Sky Survey (SDSS; originally started in 2003) contains 1.2 billion objects (about 150 GB of data). Gaia, started in 2016, delivered nearly 1.8 billion objects (about 400 GB). Looking towards the future, the Large Synoptic Survey Telescope (LSST) is projected to acquire about 1000 observations of close to 20 billion objects in 10 years of its operation (with {\em catalog} dataset sizes projected to be north of 10 PB~\citep{lsst-science_abell_2009}).

Present day survey datasets are typically delivered through a number of astronomical data archives (for example Strasbourg astronomical Data Centre, SDC \footnote{\url{http://cds.u-strasbg.fr/}}; Mikulski Archive for Space Telescopes, or MAST \footnote{\url{https://archive.stsci.edu/}}; and NASA/IPAC Infrared Science Archive, or IRSA \footnote{\url{https://irsa.ipac.caltech.edu}}), usually stored and managed in relational databases (RDBMS). The RDBMS storage allows for efficient extraction of database {\em subsets}, both horizontally (retrieving the catalog entries satisfying a certain condition) and occasionally vertically (retrieving a subset of columns).

Given their architecture and history, these systems are typically optimized for greatest performance with simple and highly selective queries. End-user analyses of archival data typically begin with subsetting from a larger catalog (i.e., SQL queries). The resulting smaller subset is downloaded to the researcher's local machine (or a small cluster). There it is analyzed, often using custom (usually Python) scripts or, more recently, code written in Jupyter notebooks \citep{ipython}. This {\em subset-download-analyze} workflow has been quite effective over the past decade, elevating surveys and archived datasets to some of the most impactful programs in astronomy. 

The upcoming increase in data volumes, as well as significant changes in the nature of scientific investigations pose challenges to the continuation of this model. The next decade is expected to be weighted towards exploratory studies examining whole datasets: performing large-scale classification (including using machine learning, ML, techniques), clustering analyses, searching for exceptional outliers, or measuring faint statistical signals. The change is driven by the needs of the big scientific questions of the day. For some — such as the nature of Dark Energy — utilization of all available data and accurate statistical treatment of measurements are the only way to make progress. For others, such as time series data, detailed insights into the variable and transient universe are now within reach from exploration of properties of {\em entire populations} of variable sources, rather than a small subset of individual objects. Secondly, the added value gained by {\em fusing information from multiple datasets} -- typically via positional cross-matching -- is now clearly understood. To give a recent example, combining information from the SDSS, Pan-STARRS, and 2MASS has recently enabled the construction of accurate maps of the distribution of stars and interstellar matter in the Milky Way \citep{3d_dust_milky_way}. Finally, a number of existing and upcoming surveys come with a {\em time domain component} -- repeat observations. These result in databases of {\em time series} of object properties ({\em light curves}, or {\em motions}), and sometime in {\em real time alerts} to changes in object properties.

All these elements present significant challenges to classical RDBMS-centric solutions, which have difficulties to scale to PB-level catalog datasets and workflows that require frequent streaming through the entire dataset. An alternative may be to consider migrating away from the traditional RDBMS-backends, and towards the management of large datasets using scalable, industry-standard, data processing frameworks built on columnar file formats that map well to the types of science analyses expected for the 2020s. Yet these systems have seen limited adoption by the broader astronomy community in large part because they lack domain-specific functionality needed for astronomical data analysis, are difficult to deploy, and/or lack ease of use.
\\

In this work, we tackle the challenge of adapting the one of the industry-standard big-data analytics frameworks -- Apache Spark -- to the needs of an astronomical end-user. We name the resultant set of changes to Spark, and the accompanying client library, {\em Astronomy eXtensions for Spark}, or AXS \footnote{The code is available at \url{https://github.com/dirac-institute/AXS} and the documentation at \url{https://dirac-institute.github.io/AXS/}}.
AXS adds astronomy-specific functionality to Spark that makes it possible to query arbitrarily large astronomical catalogs with domain-specific functionality such as positional cross-matching of multiple catalogs, region queries, and execution of complex custom data analyses over whole datasets.
With Spark's strong support for SQL, our hope is to provide a smooth migration path away from traditional RDBMS systems, and a basis for a scalable data management and analysis framework capable of supporting PB+ dataset analyses in the LSST era.

We begin by reviewing our prior work in this area and the design drivers for AXS in Section \ref{sec:prior_work}. In Section \ref{sec:spark_arch}, we introduce Apache Spark and discuss its architecture as the foundation that AXS is built on. In Section \ref{sec:crossmatch}, we describe the algorithm powering AXS distributed cross-match functionality. Section \ref{axsapi} describes the AXS Python API implementation and design choices made therein. We execute and discuss the benchmarks in Section \ref{sec:perftestresults}, and demonstrate astronomical applications on the Zwicky Transient Facility \citep[ZTF;][]{2019PASP..131a8002B, 2019arXiv190201945G} dataset in section \ref{sec:gatspy}. We summarize the conclusion and plans for future work in Section \ref{sec:conclusion}.

\section{Prior Work}
\label{sec:prior_work}

In developing the concepts for \axs, we draw heavily on experiences and lessons derived from the eight years of development, use, and support of the Large Survey Database \citep[LSD;][]{LSDurl}. 
We begin with a description of the LSD design and lessons learned from its use. We follow by discussing Spark as the basis for a next generation system, and add an overview of similar work in this area.

\subsection{The Large Survey Database}
\label{sec:prelim_lsd}

The Large Survey Database \citep[LSD;][]{2011AAS...21743319J, 2012ascl.soft09003J} is a domain-specific Python 2.7 computing framework and database management system for distributed querying, cross-matching, and analysis of large survey catalogs ($>10^9$~rows, $>1$~TB). It is optimized for fast queries and parallel scans of positionally and temporally indexed datasets. It has been shown to scale to more than $\simeq 10^2$ nodes, and was designed to be generalizable to ``shared nothing'' architectures (i.e. the ones where nodes do not share memory or disk resources and use them independently in order to maximize concurrency).  The primary driver behind its development was the immediate need for management and analysis of Pan-STARRS1 data at the Harvard-Smithsonian Center for Astrophysics, and the desire for rapid, early science from the new dataset. The LSD code is available at \url{http://github.com/mjuric/lsd}.

\subsubsection{A Spatially and Temporally Partitioned Database}

The LSD is optimized for fast queries and efficient parallel scans of positionally $(longitude, latitude)$ and temporally $(time)$ indexed sets of rows. Its design and some of the terminology have been inspired by Google's BigTable \citep{Chang:2006:BDS:1267308.1267323} distributed database and the MapReduce \citep{Dean:2004:MSD:1251254.1251264} programming model.

LSD tables are vertically partitioned into {\it column groups} ({\tt cgroups}): groups of columns with related data (e.g., astrometry, photometry, survey metadata, etc.) and horizontally into equal-area space and time cells \citep[HEALPix;][]{healpix_gorski_1998} pixels on the sky, and equal time interval. On disk, the partitioning maps to a directory structure with compressed, checksummed, HDF5 files ({\em tablets}) at the leaves. LSD achieves high performance by keeping related data physically stored together, and performing loads in large chunks (optimally $\ge128{\rm MB}$). 

\subsubsection{Fast, On-The-Fly, Positional Cross-matching}
\label{sec:lsd.xmatch}

Positional cross-match is a join of two catalogs based on the proximity of their objects' coordinates: in the simplest case, a row in table $A$ will be joined to (potentially $k$) nearest neighbors in table $B$. More usefully, the cross-match is made probabilistically, taking measurement errors and other uncertainties into account \citep{budavari16}. Cross-matching is the fundamental operation in survey data analysis. It makes it possible to associate and gather data about the same physical phenomenon (object) observed in different catalogs (wavelength regimes and timescales) or at different times within the same catalog. It allows one to query the {\em neighborhood} of the objects. Such joined and integrated information can then be fed to higher-level functions to infer the nature of the object, predict its expected behavior, and potentially assess its importance as a follow-up target (including in real time).

Though conceptually simple, spatial cross-matching is non-trivial to implement at scale over partitioned catalogs. To maximize throughput, one wishes to independently and in parallel cross-match on a per-partition (per {\em cell}, in LSD terminology) basis. If done naively, however, this could lead to incorrect cross-matches near cell edges, where the nearest neighbor resides in the adjacent partition. LSD solves the problem by duplicating in both cells rows that are within a small margin (typically, 30 arc seconds) of shared edges (the {\it neighbor cache}). This copy allows for efficient neighbor lookup (or, for example, for the application of spatial matched filters) without the need to access tablets in neighboring cells. This simple idea allows accurate cross-matches (and matched filter computations) with no inter-cell communication.

\subsubsection{Programming Model: SQL and Pipelined MapReduce}

The programming workflow model implemented in LSD is a pipelined extension of MapReduce, with an SQL-like query language used to access data. The language implemented is a subset of SQL DML \citep{ISO:2011:IIIc} -- i.e., the {\tt SELECT} statement has no subquery support -- with syntax extensions to make it easier to write astronomy-specific queries and freely mix Python and SQL. For more complex tasks, the user can write computing {\em kernels} that operate on subsets of query results on a per-cell basis, returning transformed or partially aggregated results for further processing. These are organized in a lazily computed directed acyclic graph (DAG), similar to frameworks such as Dask \citep{rocklin2015dask} or Spark\footnote{The similarity is serendipitous; when LSD was written, Spark was still in its infancy and Dask did not exist.}. Importantly, the framework takes care of their distribution, scheduling, and execution, including spill-over to disk to conserve RAM.

This combination leverages the users' familiarity with SQL, while offering a fully distributed computing environment and the ability to use Python for more complex operations. It reduced the barrier to entry to astronomers already used to accessing online archives using SQL, giving them an acceptable learning curve towards more complex distributed analyses.

\subsubsection{Science Applications}
\label{sec:sciapplications}

Initial science applications of LSD focused on managing and analyzing the $2\times10^9$ row Pan-STARRS dataset. LSD was subsequently applied to catalogs from other surveys, including the time-domain heavy PTF \citep{2009PASP..121.1395L} survey (a precursor to the ZTF), and for aspects of R\&D within the LSST Project.
\\

Notable results enabled by LSD included the next-generation dust extinction map \citep{schlafly14}, the reconstruction of the three-dimensional distribution of interstellar matter in the Galaxy from joint Pan-STARRS and SDSS observations \citep{3d_dust_milky_way, schlafly15, schlafly17}, the construction of a Gaia-PS1-SDSS (GPS1) proper motion catalog covering 3/4 of the sky \citep{tian17}, the DECam Plane Survey delivering optical photometry of two billion objects in the southern Galactic plane \citep{schlafly18}, a comprehensive map of halo substructures from the Pan-STARRS1 $3\pi$ Survey \citep{bernard16}, mapping of the asymmetric Milky Way halo and disk with Pan-STARRS and SDSS \citep[][Lurie et al., PhDT 2018]{bonaca12}, the photometric \"ubercalibration of the Pan-STARRS \citep{schlafly12}, the hypercalibration of SDSS \citep{2016ApJ...822...66F}, the analysis of the first simulated star catalog for LSST \citep[Section 4.4 in v2 onwards]{lsst-science_abell_2009}, and the discoveries and mapping of Galactic halo streams \citep{sesar2012, 2013ApJ...776...26S}.

\subsubsection{Lessons Learned with LSD}

LSD was adopted by a number of research teams as it provided them with a way to {\em query}, {\em share}, and {\em analyze} large datasets in a way responsive to their customs and needs. It reduced the barrier to entry for analysis from having the skill to write a complex SQL query (with potentially multiple subqueries and table-valued function tricks) to having the know-how to write a Python function operating in parallel on chunks of the result sets. Importantly, it was {\em performant} and {\em transparent}. LSD dataset scans typically finished in a fraction of time required for analogous queries in the central RDBMS solution (running low-selectivity queries incurred virtually no overhead over raw {\tt fread}-type I/O). By providing a clear and simple programming model, it avoided the downside of complex query optimizers that sometimes make innocuous changes to queries and lead to orders of magnitude different performance, frustrating users.

Secondly, LSD served as a test-bed for new concepts and technologies, validating design choices such as column-store data structures, Python-driven distributed workflows organized in DAGs, aggressive in-memory operation with in- and out-of-core caching, mixing of declarative (SQL) and imperative syntax and others. When initially implemented in LSD, these were experimental and controversial choices; today they're broadly used and robustly implemented by frameworks like Spark, Dask, and Pandas and formats like Parquet \citep{Parquet}.
\\

\noindent Just as importantly, the years of ``in-the-field'' experience gives us an opportunity to understand the major areas in need of improvement:
\begin{itemize}
	\item {\bf Fixed partitioning:} LSD implements a fixed, non-hierarchical, partitioning scheme\footnote{A conscious decision to keep the design simple, and accelerate early development.}. This leads to significant partitioning skew (factors of 100 between the rarest and densest areas of the sky). While partitioning can be changed on a per-table basis, cross-matching tables with different partitionings is not possible.

	\item {\bf Problematic temporal partitioning:} LSD tables are partitioned on time, facilitating performant appends and ``time slicing''. However, the vast majority of real-world use-cases have users request {\em all} the data on a particular object (typically to perform classification or some other inference task). Having the time series scattered over a large number of files induces a significant performance hit. A design where time-series is stored as an array column within the object table would perform significantly better.

	\item {\bf Lack of robust distributed functionality:} LSD is not resilient to the failures of workers and/or processing nodes. Setting it up and running in multi-node configurations was always experimental and a major end-user challenge. This functionality is crucial for it to continue to scale, however.

	\item {\bf Much custom code and no Python 3 support:} LSD contains custom code and solutions for problems where mature, adopted, solutions exist today (Pandas, AstroPy, scikit-learn, and Spark itself). This reduces stability, developer community buy-in, and increases maintenance cost. Also, LSD is written in Python 2.7, for which support will end in 2020.
\end{itemize}

\subsection{Desiderata for a next-generation system}

The issues identified in the previous section, and especially the point about custom code, made us reexamine the development path for LSD. Rather than continuing to maintain an old (and in many ways experimental) code base, a more sustainable way forward would be to build a new system that retains the successful architectural concepts and the user-friendly spirit of LSD while addressing the recognized issues. In particular, we define the following set of key desiderata:
\begin{enumerate}
    \item {\bf Astronomy-specific operations support}: for an analysis system to be broadly useful to astronomers it must support common astronomy-specific operations, the most important of which are cross-matching and spatial querying,
    \item {\bf Time-series awareness}: the ability to intuitively query or manipulate entire time series of observations of a single object or an entire population,
    \item {\bf Ease of use}: enable the domain scientist to construct and execute arbitrary analyses, {\em in a distributed fashion} by mixing declarative SQL syntax and Python code, as appropriate,
    \item {\bf Efficiency}: fast execution of key operations (positional cross-matching, selective filtering, and scanning through the entire dataset),
    \item {\bf Scalability}: ability to handle O(10TB) and scale to O(1PB+) tabular datasets, with significant data skews,
    \item {\bf Use of industry-standard frameworks and libraries}: building on present-day, proven, technologies makes the code more maintainable in the long-run. It also allows us to leverage the R\&D, code, and services, from other areas of big data analyses.
\end{enumerate}

This last element -- the maximal re-use of industry standard frameworks -- has been a particularly strong driver in this work. Relative to other similar approaches (e.g., \citet{astroide_brahem_2018}; see also Section~\ref{sec:others}), we aim to build {\bf on top} of Apache Spark, with minimal changes to the underlying storage scheme, engine, or query optimizer. We therefore design our changes and algorithms to make as many of them {\em generic}, and admissible for merging into the Apache Spark mainline. While this restricts our choices somewhat, it enables (the much larger) communities outside astronomy to benefit from our improvements, as well as contribute to them. It also increases long-term maintainability by reducing the astronomy-specific codebase needed to be maintained by the (much smaller) astronomy community.

\section{Astronomy eXtensions for Spark}
\label{sec:spark_arch}
\subsection{Apache Spark as a basis for Astronomical Data Management and Analysis Systems}
\label{sec:org19e8774}

Apache Spark is a fast and general-purpose engine for big data processing, with built-in modules for streaming, SQL queries, machine learning, and graph processing \citep{spark_zaharia_2016, armbrust2015}. Originally developed at UC Berkeley in 2009, it has become the dominant big data processing engine due to its speed (10-100x faster than Hadoop), attention to ease of use, and strong cross-language support. Spark supports performant column-store storage formats such as Parquet \citep{Parquet}. It is scalable, resilient to individual worker failures and provides strong Python interfaces familiar to astrophysicists and other data scientists. Similar to LSD, Spark already ``... offers much tighter integration between relational and procedural processing, through a declarative DataFrame API that integrates with procedural code'' \citep{armbrust2015}. Relative to comparable projects such as Dask \citep{rocklin2015dask}, Spark is more mature and broadly adopted. Furthermore, Spark natively supports {\em streaming} (and with a unified DataFrames API), allowing applications to real-time use cases.

\subsection{Spark Architecture}

Fundamentally, Spark is a system that facilitates fault-tolerant, distributed, {\em transformations} of arbitrarily large datasets\footnote{E.g., even an SQL query can be thought of as transforming the original, potentially PB-scale dataset, to a new, potentially few-kB dataset.}. The core Spark abstraction is one of a {\em resilient distributed dataset}, or RDD, a fault-tolerant collection of elements (e.g., table rows, images, or other datums) that can be operated on in parallel. Parallelism is achieved through {\em partitioning}: each RDD is composed of a number of (potentially large) {\em partitions}, with transformations operating primarily on individual partitions. Spark provides a number of fundamental transformations, such as {\em map} (where each element in the input dataset is operated on by an arbitrary function to be mapped into the output dataset), {\em filter} (where only elements for which the filtering function returns true are placed into the output dataset), {\em reduceByKey} (where a reduction operation is applied to all elements with the same key), and others\footnote{E.g., see \url{https://spark.apache.org/docs/latest/rdd-programming-guide.html\#transformations} for a complete list}. Any non-fatal failures (e.g., temporary network outages, or nodes down) are detected automatically, and computuation is re-scheduled and re-executed with no need for user involvement.

Nearly every data analysis algorithm can be expressed as a set of transformations and actions on Spark RDDs. This fact is used to build higher-level abstractions that retain the distributed and fault-tolerant properties of RDDs, but provide a friendlier (or just more familiar) interface to the user. Of particular importance for us are the {\em Spark SQL} layer and the {\em DataFrames} abstraction.

Spark SQL implements handling of structured data organized in typed columns -- tables -- using extended standard SQL language and the matching API. This enables one to write SQL queries over any tabular dataset stored in a format that Spark is able to read. The core data structure Spark SQL operates on is the \texttt{DataFrame} (a generalized table). Spark DataFrames are built on RDDs, inheriting their fault tolerance and distribution properties. Otherwise, they are very similar in behavior (and API) to R or Pandas data frames. Beyond a friendlier API, DataFrames enable performance optimizations provided by Spark's Catalyst optimizer: advanced query planning and optimization, translation of SQL queries and transformations into generated Java code compiled and executed on the fly, and a more compact data serialization.

On top of this structure lies a rich set of libraries, used by end-user application programs. For example, Spark's MLlib includes efficient distributed implementations of major machine learning algorithms. This layer also includes various programming language bindings, including Python. 
All these characteristics make it an excellent choice as a basis for an end-user framework for querying, analyzing and processing catalogs from large astronomical surveys. They tick off nearly all technical desiderata (Efficiency, Scalability, and Broad Support), leaving us to focus on the addition of astronomy-specific operations, time-series awareness, and the ease of use through an enhanced Python API. We describe these in the sections to follow.

\begin{figure}
	\begin{center}
		\includegraphics[width=0.48\textwidth]{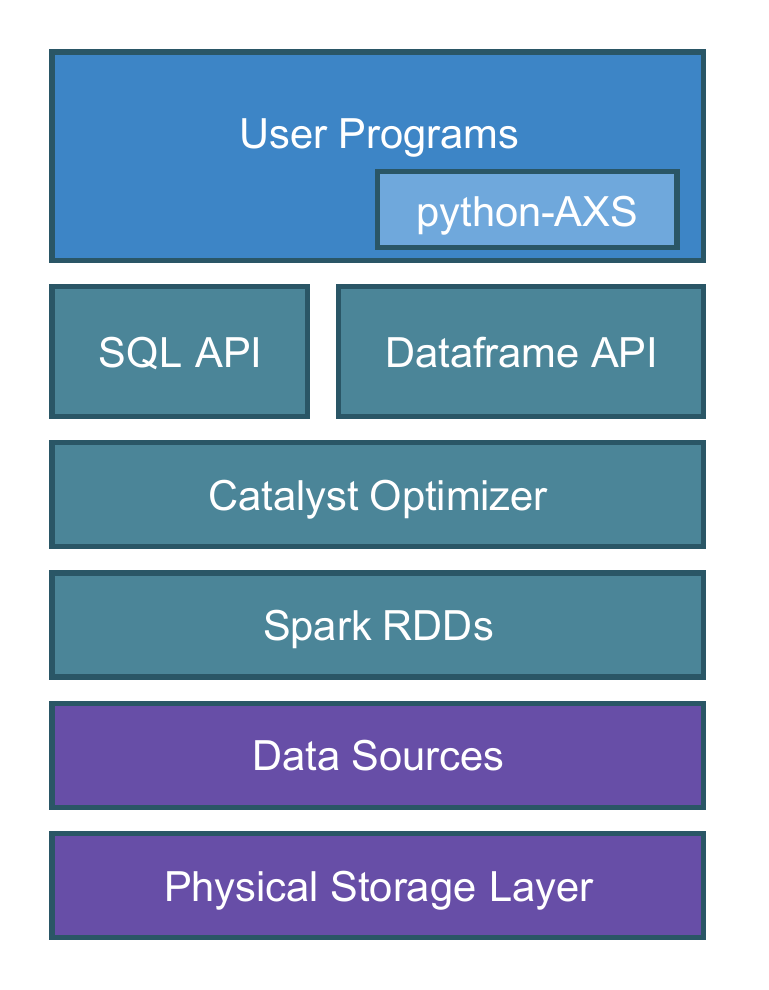}
	\end{center}
	\caption{The layered architecture of Spark, with the Python elements of \axs. By the end of the grant period, our entire code will reside in the Python layer. The modifications to Data Sources, RDD, Catalyst and DataFrame layers will be submitted to upstream projects enabling their broader use.\\ }
	\label{fig:sparkArch}
\end{figure}

\section{Extending Spark SQL Library to Enable Astronomy-Specific Operations}
\label{sec:crossmatch}

To support an efficient and scalable cross-match and region queries of data stored in Spark DataFrames, we minimally extend Spark API with two main contributions: a data partitioning and distribution scheme, and a generic optimization of Spark's sort-merge join algorithm on top of which we build a distributed positional cross-match algorithm.

\subsection{Cross-matching objects in astronomical catalogs}
\label{sec:orgd4bf806}
In astronomy, we are often interested in joining observations from two (or more) survey catalogs that correspond to the same physical objects in the sky. The simplest version of this problem reduces to finding all observations, from two or more catalogs, that are less than some angular distance apart.

More formally, if $L$ is the left relation (catalog) and $R$ is the right one, the cross-matching operation is a set of pairs of tuples $l$ and $r$ such that the distance between them is less than the defined threshold of $\epsilon$:

\begin{equation}
    \{ (l, r) \; | \; (l, r) \in L \times R, dist (l, r) \leq \varepsilon \}
\end{equation}

\begin{figure}[htbp]
\centering
\includegraphics[width=.9\linewidth]{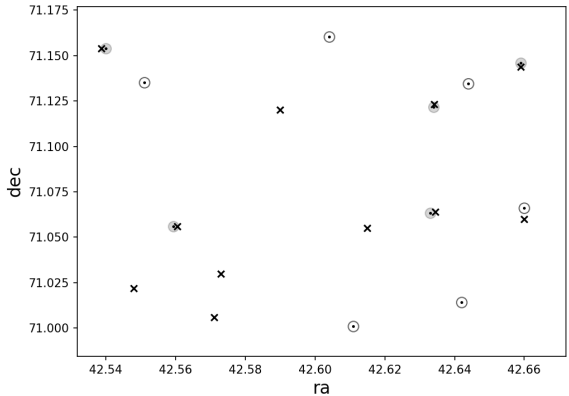}
\caption{\label{fig:org89e7d56}
An example of cross-matching two catalogs in a 10 arcmin x 10 arcmin region of the sky. Objects of the two catalogs being matched are represented as dots and crosses. The circles show the search region around each object of the first catalog. Circles are filled if their area contains a match.}
\end{figure}

This is graphically illustrated in Figure \ref{fig:org89e7d56}.

AXS also supports the nearest-neighbor join where, for each tuple in the left relation, only the tuple in the right relation with the minimum distance is included (which still has to be smaller than the defined $\epsilon$ threshold). 

\subsection{Distributed zones algorithm}
\label{sec:distributed}

The fundamental problem with cross-matching is how to enable data spatial locality and organize the data for quick searches. Traditionally, two different indexing schemes have been popular: HEALPix (Hierarchical Equal Area and isoLatitutde Pixelization of the sphere)~\citep{healpix_gorski_1998} and HTM (Hierarchical Triangular Mesh)~\citep{htm_kunszt_2001}. In one of the earlier papers~\citep{there-goes_gray_2004} Gray et al. compare HTM, which is used in SDSS' SkyServer~\citep{data-mining_gray_2002}, and the zones approach and find the zones indexing scheme to be more efficient when implemented within relational databases. The zones algorithm is further developed in~\cite{zones_gray_2007}.

For AXS, we extend the \citet{zones_gray_2007} Zones algorithm to adapt it for a distributed, shared-nothing architecture. Zones divides the sky into horizontal stripes called "zones" which serve as indexes into subsets of catalog data so as to reduce the amount of data that needs to be searched for potential matches. Given such partitioning, the general idea is to express the cross-join operation as a query of the form shown in Listing~\ref{query:join}.

\begin{lstlisting}[caption=Example of a range query, label=query:join]
SELECT * from GAIA g JOIN SDSS s 
  ON g.zone = s.zone 
  AND g.ra BETWEEN s.ra - e AND s.ra + e
  AND distance(g.ra, g.dec, s.ra, 
    s.dec) <= e
\end{lstlisting}
(where \texttt{e} is a distance that defines the size of the moving window and \texttt{distance} is a function which calculates distance between two points), but ensure that a) the data is partitioned so that this query can be run in parallel over many partitions, and b) the SparkSQL optimizer is capable optimizing the query so as to avoid a full cartesian JOIN within each partition (i.e., that the expensive \texttt{distance} function is evaluated only on pairs of objects within a bounding box defined by the \texttt{g.ra BETWEEN s.ra - e AND s.ra + e} clause of the query).

\subsubsection{Bucketing using Parquet files\label{parquet}}
In the distributed zones algorithm we keep the division of sky into N zones but also physically partition data into B buckets, implemented as Parquet bucketed files, so as to enable independent and parallel processing of different zones. Parquet\footnote{https://parquet.apache.org/} is a distributed and columnar storage format with compression and basic indexing capabilities. Parquet's columnar design is a major advantage for most astronomical catalog usage.
Buckets in a Parquet file are actually separate physical Parquet files in a common folder. The Parquet API then handles the whole collection of files as a single Parquet file. Buckets could be implemented in different ways, but we chose Parquet because of its ease of use, level of integration with Spark and performance.

\begin{figure}[htbp]
\centering
\includegraphics[width=.9\linewidth]{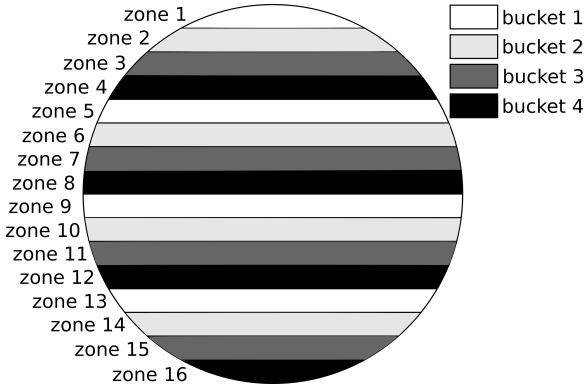}
\caption{\label{fig:org29d704b}
Partitioning the sky into zones and placing zones into buckets. The example shows the sky partitioned into 16 horizontal zones. Objects from each zone get placed into buckets sequentially. In reality, zones are much narrower and are counted in thousands.}
\end{figure}

All the objects from the same zone end up in the same bucket. The zones are placed in buckets sequentially: zone \texttt{z} is placed into bucket \texttt{b = z \% B} (figure \ref{fig:org29d704b} shows an example for 16 zones and 4 buckets, but in reality thousands of zones are used). The reason for placing zones into buckets in this manner is that placing thin neighboring stripes into different buckets automatically \textit{reduces data skew}, so often present in astronomical catalogs.

If we partition two catalogs in the same way (with the same zone height and the same number of buckets), we can cross-match objects within the same buckets independently of the other buckets. This scheme makes the cross-join operation parallelizable and scalable.

\subsubsection{Joining data within buckets}

Data in each bucket are sorted first by \texttt{zone}, then \texttt{ra} column, which serve as indexing columns for cross-matching operations. Every table handled by AXS needs to have \texttt{zone} and \texttt{ra} columns if it is to be used for subsequent cross-matching with other tables (also, zone height and number of buckets in the two tables being joined need to be the same).

In order to efficiently join data within buckets, we chose to use an "epsilon join"~\citep{epsilon_silva_2010} implementation where two tables are joined quickly in one pass by maintaining a moving window over the data in the right table, based on an equi-join condition on the primary column (\texttt{zone} in our case) and a range condition on the secondary column (\texttt{ra} in our case). Importantly, we extended the Spark's sort-merge join implementation~\footnote{https://github.com/apache/spark/pull/21109} to recognize an optimization opportunity and avoid calculating the distance function for all object pairs from the two zones, but only for those which match the prior \texttt{BETWEEN} condition.

\begin{figure}[htbp]
\centering
\includegraphics[width=.9\linewidth]{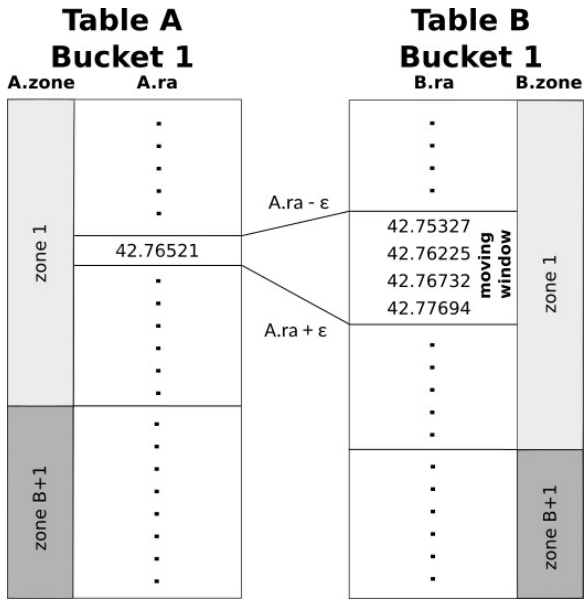}
\caption{\label{fig:org145de84}
Using the "epsilon join" to reduce the number of rows for which distance is calculated. For the match candidate row in the figure, only four distance calculations are performed. (B stands for the number of buckets.)}
\end{figure}

As a consequence of data bucketing scheme and the epsilon join optimization, Spark executes a query shown in Listing~\ref{query:join} as multiple parallel and fast join operations on bucket pairs from the two tables. The join is performed without data exchanges (shuffles) between nodes, and the data needs to be read only once. Figure \ref{fig:org145de84} shows the process for a bucket pair graphically.

\subsection{Correctness at zone boundaries}
\label{sec:orga81a592}
An important point for cross-match completeness and performance is joining data from neighboring zones. Objects residing near zone borders might have matching objects in the border area of the neighboring zone. Joining only objects from the matching zones would miss these matches in the neighboring zones. In order to maintain the possibility to join objects within a single zone independently of other zones, we chose to duplicate objects from the lower border stripe of each zone to the zone below it. This marginally increases the amount of data stored on disk, but allows us to run cross-joins without data movement between processes (and possibly nodes) during query processing.

The duplication does have consequences for ordinary table queries. First, queries for objects in the catalog now need to be modified so as not to include the duplicated data. Second, cross-matching results must be pruned for objects from duplicated border stripes that have been cross-matched twice (once inside their original zone and once inside the neighboring zone). In our implementation, this is transparently handled in the Python AXS API layer.

Finally, we note that the ``height`` of stripe overlap needs to be selected such that it covers the maximum expected cross-join search radius. We find that $10 {\rm arcsec}$ is a reasonable default, capable of handling datasets ranging from SDSS to ZTF and Gaia scales.

\subsection{The zone height and the number of buckets}
\label{zoneheightbuckets}
Choosing the number of buckets is a trade-off between the bucket size (and the amount of data that needs to be read as part of a single Spark task) and the maximum parallelism achievable. This is because a single task can process all buckets serially, but several tasks cannot process a single bucket. Data is read from buckets in a streaming fashion, so larger buckets do not place much larger burden on memory requirements.

Larger zones reduce the number of zones which means that less data will need to be duplicated. However, smaller zones reduce data skew. Data skew can significantly affect processing, so it is advisable not to use zones that are too large. Furthermore, larger zones will have more rows in their ``moving windows'' during the cross-match operation and will hence require more memory for cross-matching.

\subsection{ Data skew considerations }
\label{sec:org0005f3e}
Large (especially all-sky) astronomical datasets often include highly skewed data because of the highly uneven distribution of astronomical sources on the sky. Under such skew and with naive spatial partitioning, processing of queries on very populous partitions would use large amounts of memory and last much longer than what would typically be the case. This would considerably degrade the overall query performance. 

As was already mentioned in section \ref{parquet}, a convenient property of data placement according to the distributed zones algorithm is that, when zones are sufficiently narrow, the data gets naturally distributed more or less equally among the buckets. In practice, even for highly skewed catalogs such as SDSS and Gaia, we observed the maximum ratio of sizes of different bucket files of a factor of two.

\section{Ease of Use: AXS Python API}
\label{axsapi}
As the main end-user interface to AXS we provide a Python API. This API is a thin layer on top of {\tt pyspark}, Sparks's own Python API. We designed it to expose the added astronomy-specific functionality, while abstracting away the implementation details such as partitioning or the underlying cross-match algorithm.

In this section we highlight some of the key elements of the Python API. The full documentation is available at \url{https://dirac-institute.github.io/AXS/}.

\subsection{A Simple Example}
\label{sec:apiexample}
The two main classess the user interacts with in AXS are \texttt{AxsCatalog} and \texttt{AxsFrame}. These serve as extensions of Spark's \texttt{Catalog} and \texttt{DataFrame} interfaces, respectively.

By analogy with Spark's {\tt Catalog}, an instance of \texttt{AxsCatalog} is constructed from an instance of \texttt{SparkSession}:
\begin{lstlisting}[language=Python]
from axs import AxsCatalog
axs_catalog = AxsCatalog(spark)
\end{lstlisting}
The \texttt{SparkSession} object is similar to a connection object from standard Python database API\footnote{\url{https://www.python.org/dev/peps/pep-0249/}}. It represents the connection to the Spark metastore database, and enables manipulation of Spark tables. The \texttt{AxsCatalog} instance adds awareness of tables that have been partitioned using the distributed zones algorithm (Section \ref{sec:distributed}), and the ability to retrieve their instances as shown in the following snippet:
\begin{lstlisting}[language=Python]
axs_catalog.list_tables()  # output omitted
sdss = axs_catalog.load("sdss")
gaia = axs_catalog.load("gaia")
\end{lstlisting}

The returned objects above are \texttt{AxsFrame}s. These extend Spark \texttt{DataFrame} objects with astronomy-specific methods. One of these is \texttt{crossmatch} which performs cross-matching between two \texttt{AxsFrame} tables. 
\begin{lstlisting}[language=Python]
from axs import Constants
gaia_sd_cross = gaia.crossmatch(sdss, 
  r=3*Constants.ONE_ASEC, return_min=False)
gaia_sd_cross.select(
  "ra", "dec", "g", "phot_g_mean_mag").
  save_axs_table("gaiaSdssMagnitudes")
\end{lstlisting}

The snippet above sets up a pipeline for positional cross-matching of the {\tt gaia} and {\tt sdss} catalogs. The resulting catalog is then queried for a subset of columns, which are finally saved into a new, ZONES-partitioned, table named {\tt gaiaSdssMagnitudes}. We note that the graph above executes only when {\tt save\_axs\_table} is called; in other words, {\tt crossmatch} and {\tt select} are {\em transformations} while {\tt save\_axs\_table} is an {\em action}\footnote{See \url{https://spark.apache.org/docs/latest/rdd-programming-guide.html\#rdd-operations} for details.}.

Had the {\tt return\_min} flag in the above snippet been set to {\tt True}, {\tt crossmatch} would return only the nearest neighbor for each row in the Gaia catalog. 

\subsection{Support for Spatial Selection}
\label{sec:orgf1cc66a}
{\em Region queries} in AXS are queries of objects in regions of the sky based on boundaries expressed as minimum and maximum RA and DEC angles. Having sky partitioned in zones and zones stored in buckets, region queries get translated into several searches through the matching bucket files. The underlying Spark engine is able to execute these in parallel. And since the bucket files are sorted by \texttt{zone} and \texttt{ra} columns, the zone being derived from DEC values, these searches can be performed quickly. We take advantage of the fact that Spark can push column filters down to Parquet file reading processes so that whole bucketed files can be skipped if they are not needed. Similarly, reading processes can skip parts of files, based on sort columns. 

These optimizations are hidden from the AXS user and region queries in AXS API are as simple as this:
\begin{lstlisting}
region_df = gaia.region(ra1=40, dec1=15, 
                        ra2=41, dec2=16)
\end{lstlisting}

Cone search is implemented using the \texttt{cone} method. It requires a center point and a radius and returns all objects with coordinates within the circle thus defined.

More complex selections (e.g., support for polygons) are possible to implement using these primitives coupled with additional Python code. These may be added at a later date, if there's sufficient user demand.

\subsection{Support for Time Series}
\label{sec:org34773cc}
Support for time series data that is both performant and user friendly is increasingly important given the advent of large-scale time-domain surveys. A prototypical time series is a light curve: a series of flux (magnitude) measurements of a single object. Measurements typically include the measurement itself, the time of the measurement, the error, and possibly other metadata (e.g., flags, or the filter band in which the measurement was taken).

With AXS, we recommend storing time series data {\em as set of vector (array) columns in an object catalog}. This is a departure from the more classical RDBMS layout employing an ``object'' and ``observation'' table, where the light curve would be constructed by {\tt JOIN}-ing the two. The classical approach has two disadvantages: converting the returned JOIN into a more natural ``object + light curve'' representation is left to the user, and may not be trivial (especially to an inexperienced user). Secondly, the need to perform a JOIN, and the fact that the observation data is physically separate (on disk) from object data, reduces the overall query performance. The downside of our approach is that updates to time-series columns are expensive; this, however, is not an issue when AXS is used to manipulate large {\em data release} datasets which are static by definition. Finally, nothing precludes one from structuring their dataset across two (object, observation) tables if it's better for their particular use case. Various time-series support functions assume the vector-column format at this time, however.

We illustrate a few time-series support functions on an example of handling a light curve. To make it simpler to support multi-wavelength use cases (where the light-curve across all bands is passed to a user-defined function for, e.g., classification), we recommend storing observations in a tuple of vector columns such as {\tt (time, mag, magErr, band)}. This does reduce performance if the user wishes to perform analysis on a single filter at a time, but we increasingly see a demand for simultaneous cross-band analysis. To support maximum performance, we provide two helper array functions usable from within queries: \texttt{ARRAY\_ALLPOSITIONS}, which returns an array of indexes of all occurrences of an element, and \texttt{ARRAY\_SELECT}, which returns all elements indexed by the provided index array. These two functions, combined with other Spark SQL functions, can be used for querying and manipulating light-curve data in AXS tables.

For example, to get the number of r-band observations for all objects in a catalog {\tt ztf}, assuming that column \texttt{band} contains the filter name used for each measurement, one can use a snippet similar to the one in Listing~\ref{py:array}.
\begin{lstlisting}[caption=Example of using {\tt array\_allpositions}, label=py:array]
from pyspark.sql.functions import size, 
  array_allpositions
ztf_rno = ztf.select(size(
  array_allpositions(ztf("band"), "r")))
\end{lstlisting}
\texttt{array\_allpositions} returns an array of indices into the \textit{band} column, each corresponding to the \textit{r}-band value. The Spark's built-in function \texttt{size} returns the length of the indices array. 

\subsection{Fast Histograms}
\label{sec:orga39b57a}
A common summarization technique with large survey datasets is to build {\em histograms} of a statistic as a function of some parameters of interest. Examples include sky maps (e.g., counts or metallicity as a function of on-sky coordinates) and Hess diagrams (counts as a function of color and apparent magnitude). AXS Python API offers a support for straightforward creation of histograms. These are implemented as relatively straightforward wrappers around Spark API, making their execution fully distributed and fault-tolerant. 

The functions of interest are {\tt histogram} and {\tt histogra- m2d}. When calling the \texttt{histogram} method, users pass a column and a number of bins into which the data is to be summarized. Similarly, \texttt{histogram2d(cond1, cond2, numbins1, numbins2)} bins the data in two dimensions, using the two provided condition expressions.

An example of histogramming is given in code Listing~\ref{py:hist2d}. The code results in a 2D histogram graph showing the density of differences in observations in {\it g band} between SDSS and Gaia catalogs, versus the same differences between WISE and Gaia catalogs. Both differences are binned into 100 bins.

\begin{lstlisting}[caption=An example of using \texttt{histogram2d}, label=py:hist2d]
from pyspark.sql.functions import coalesce
import matplotlib.pyplot as plt
cm = gaia.crossmatch(sdss).crossmatch(wise)
(x, y, z) = cm.histogram2d(
  cm.g - cm.phot_g_mean_mag, 
  cm.w1mag - cm.phot_g_mean_mag, 
  100, 100)
plt.pcolormesh(x, y, z)
\end{lstlisting}

\subsection{Saving intermediates}
\label{sec:orgfd7a91d}
By default, Spark eschewes saving intermediate results of computations, unless an explicit request has been made to do so.

We've found there are common use cases where saving intermediate calculation results and reusing them later is useful. One example is a result of a cross-match of large tables, which will then be further joined with or cross-matched to other catalogs. To be subsequently cross-matched, these intermediate tables need to be partitioned in the same distributed ZONES format and stored in the {\tt AxsCatalog} registry. To support this common operation, we provide an \texttt{AxsFrame.save\_axs\_table} method. It saves the underlying \texttt{AxsFrame}'s data as a new table and partitions the data as was described in section \ref{sec:distributed}.

\subsection{Support for Python User-Defined Functions (UDFs)}
\texttt{AxsFrame} class' methods \texttt{add\_column} and \\
\texttt{add\_primitive\_column} are thin wrappers around Spark's \texttt{pandas\_udf} and \texttt{udf} functions. They are only intended to make it a bit easier for astronomers to run custom data processing functions on a row-by-row basis (i.e. to avoid using \texttt{@pandas\_udf} and \texttt{@udf} annotations) and make their code more readable. They are applicable only when handling data row by row, which corresponds to Spark's \texttt{udf} function and Spark's \texttt{pandas\_udf} function of type \texttt{PandasUDFType.SCALAR} and cannot be used for \texttt{PandasUDFType.GROUPED\_MAP} nor \texttt{PandasUDFType.GROUPED\_AGG} UDFs. 

Both functions accept a name and a type of the column to be added, the function to be used for calculating the column's contents, and names of columns whose contents are to be supplied as input to the provided function. The difference between the two methods is that \texttt{add\_primitive\_column} supports only outputting columns of primitive types, but is significantly faster because it uses Spark's \texttt{pandas\_udf} support under the hood. \texttt{add\_column} method uses the scalar \texttt{udf} functions, making it slower, but supports columns of complex types. \texttt{pandas\_udf} is faster because it is able to handle blocks of rows at once by utilizing Python Pandas framework (and its vectorized processing).

For an example and discussion of using UDFs see Section~\ref{sec:gatspy}.

\subsection{Adding New Data to AXS Catalogs}
\label{sec:orgd6a7f65}
While our primary use-case at present is to support analysis of large, static (i.e., data release) datasets, we've found it useful to be able to incrementally add data to AXS tables (e.g., to facilitate incremental ingestion).

For this purpose, AXS provides the method \\
\texttt{add\_increment}:
\begin{verbatim}
add_increment(self, table_name, increment_df, 
    rename_to=None, temp_tbl_name=None)
\end{verbatim}

This will add the contents of the \texttt{increment\_df} Spark DataFrame to the AXS table with the name \texttt{table\_name}, taking care to calculate zones, and to bucket and sort the data appropriately. Users can customize the table name for the copy of the old data (\texttt{rename\_to}) and the temporary table name used in the process (\texttt{temp\_tbl\_name}).

Adding an increment to an existing table means that the new and old data need to be merged, repartitioned, and saved as a separate table. On top of physical movement of data during this operation, time has to be spent on data sorting and partitioning, which makes this operation the most expensive of all the operations presented so far. Data partitioning and sorting is a necessary part of the cross-match operation, but is performed in advance and only once, so that the latter part (table joins) can be performed online as needed. 

We measured the time AXS needs to partition the different catalogs. You can find the results in section \ref{partperf}.

\section{Cross-matching performance}
\label{sec:perftestresults}

To test AXS' cross-matching performance, we used the catalogs listed in Table \ref{tblcatalogs}. We tested both the scenario where all matches within the defined radius were returned, and the scenario where only the first nearest neighbor was returned for each row. The number of resulting rows for each catalog cross-match combination is given in Table \ref{tblcms}, for both scenarios. Furthermore, we compared cross-matching performance for both scenarios in cases when the data was cached in the OS buffers (warm cache) and when the data wasn't cached (the cache was empty, or "cold cache") \footnote{For clearing the OS buffers and creating a "cold cache" situation we wrote "3" to the \textit{/proc/sys/vm/drop\_caches} file}. This caching mechanism works on OS level and is separate from Spark's caching mechanism (which we didn't use) and from Spark's memory handling. 

\begin{figure*}[htbp]
\centering
\includegraphics[width=\textwidth]{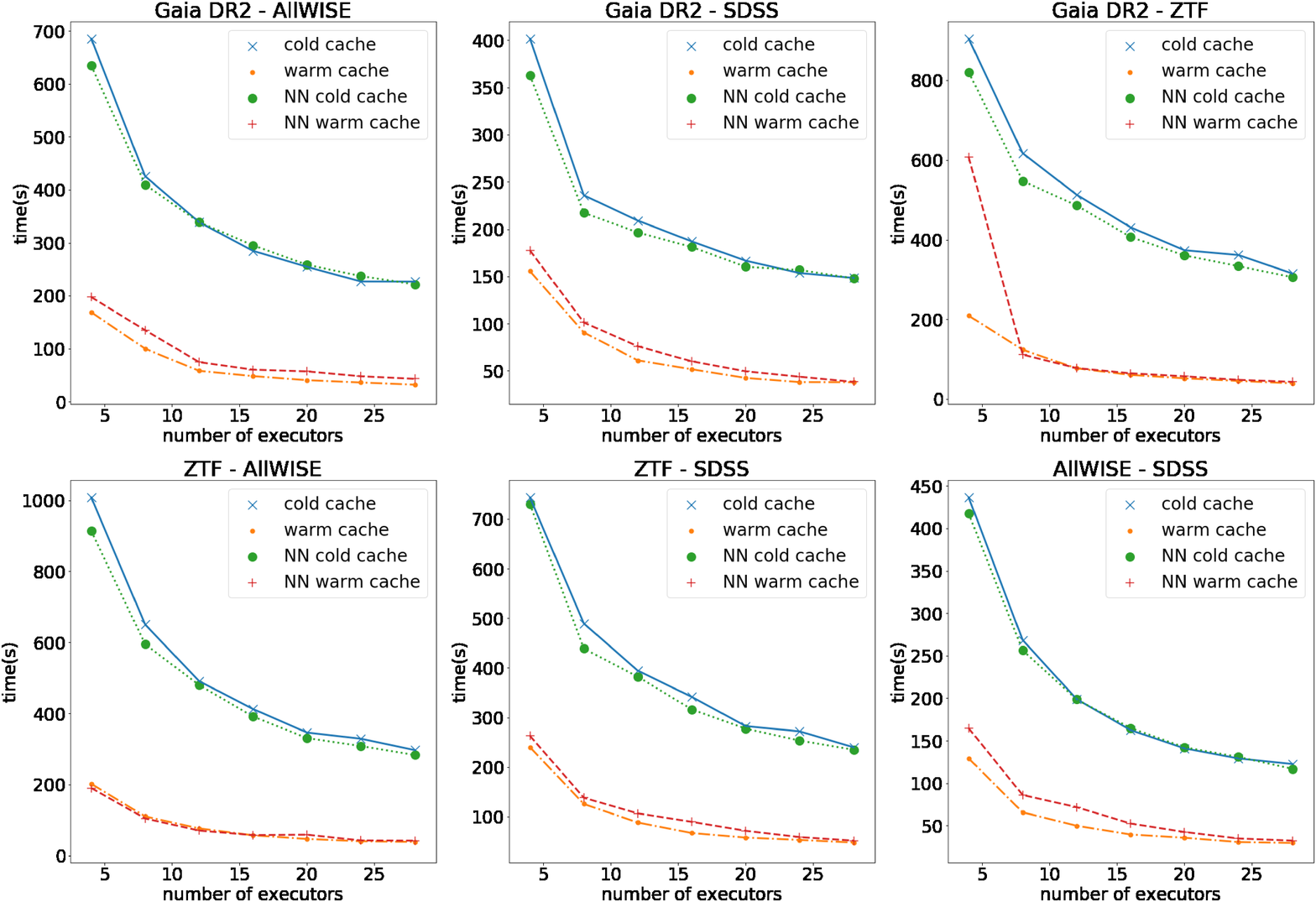}
\caption{\label{figcmboth}
Performance tests of cross-matching various catalogs in scenarios with file system buffers empty or full (we used Linux OS-level caching, not Spark caching), when returning all matches or just the first nearest neighbor ("NN" results).}
\end{figure*}

For each scenario we also varied the number of Spark executors (level of parallelism) in the cluster (going from 1 to 28). Each executor was given 12 GB of Java memory heap and 12 GB of off-heap memory. \footnote{The ability to directly allocate memory off Java heap was introduced into Spark as part of Tungsten project's changes and can considerably improve performance by avoiding Java garbage collection.} The results are shown in figure \ref{figcmboth}, and the raw results are in the tables Table \ref{tblcmraw1} and Table \ref{tblcmraw2} for the first scenario, and tables Table \ref{tblcmrawg1} and Table \ref{tblcmrawg2} for the second scenario. Each data point in the figure and in the tables is an average of three tests.

The results demonstrate the performance that can be expected when running AXS on a single machine. The scalability is constrained by the single machine's shared resources and, consequently, the performance doesn't improve much beyond 28 executors. It should also be noted that for these tests we used a local file system, not HDFS (tests in a distributed environment using data from S3 storage and HDFS are planned for the future and will hopefully be published in a subsequent paper).

The warm cache results show the ``raw'' performance of the cross-matching algorithm, not including the time required for reading the data from disk. It should be noted, however, that the cache didn't have the capacity for complete datasets and that data was partly read from disk in all cases. The cold cache results show the performance users can expect if they don't have much memory available. 

These cross-matching results outperform other systems, although direct comparisons are difficult because of different architectures, datasets, and algorithms used. Other teams report best cross-matching times on the scale of tens of minutes, while our best results are in tens of seconds. The comparisons are discussed in Section~\ref{sec:others}. 

\begin{table}
    \begin{tabular}{ l | r | r | r }
        Catalog & Row count & R. cnt. no dup. & Size \\
        \hline
         SDSS & 0.83 Bn & 0.71 Bn & 71 GB \\
         Gaia DR2 & 1.98 Bn & 1.69 Bn & 464 GB \\
         AllWISe & 0.87 Bn & 0.75 Bn & 384 GB \\
         ZTF & 3.13 Bn & 2.93 Bn & 1.17 TB \\
    \end{tabular}
    \caption{ The catalog used for performance tests, with the number of rows, number of non-duplicated rows and compressed data size. }
    \label{tblcatalogs}
\end{table}

\begin{table}
    \begin{tabular}{ l | l | r | r }
        Left cat. & Right cat. & Results - all & Results - NN \\
        \hline
         Gaia DR2 & AllWISE & 320 M & 320 M \\
         Gaia DR2 & SDSS & 227 M & 126 M \\
         ZTF & AllWISE & 109 M & 109 M \\
         ZTF & SDSS & 273 M & 168 M \\
         Gaia DR2 & ZTF & 92 M & 49 M \\
         AllWISE & SDSS & 235 M & 119 M \\
    \end{tabular}
    \caption{ List of catalog combinations used for cross-match performance tests, with the numbers of resulting rows when returning all matches or only the first nearest neighbor. }
    \label{tblcms}
\end{table}

\begin{table}
    \begin{tabular}{ c | c | c | c | c | c | c }
        \multirow{2}{*}{Execs.} & \multicolumn{2}{c|}{Gaia-AllWISE} & \multicolumn{2}{c|}{Gaia-SDSS}  & \multicolumn{2}{c}{ZTF-AllWISE} \\
        \cline{2-7}
        & warm & cold & warm & cold & warm & cold \\
        \hline
        1 & 2438 & 2276 & 1441 & 1412 & 3487 & 3283 \\
        2 & 313 & 1276 & 300 & 744 & 1772 & 1732 \\
        4 & 168 & 685 & 155 & 401 & 202 & 1007 \\
        8 & 99 & 425 & 90 & 235 & 110 & 650 \\
        12 & 58 & 338 & 61 & 209 & 76 & 491 \\
        16 & 48 & 284 & 51 & 187 & 57 & 412 \\
        20 & 40 & 254 & 42 & 166 & 47 & 346 \\
        24 & 35 & 226 & 37 & 153 & 40 & 329 \\
        28 & 31 & 226 & 37 & 148 & 38 & 296
    \end{tabular}
    \caption{ Averaged raw cross-match performance results (in seconds), when returning all matches, for the first three catalog combinations, depending on the number of executors and whether cold or warm OS cache was used. }
    \label{tblcmraw1}
\end{table}

\begin{table}
    \begin{tabular}{ c | c | c | c | c | c | c }
        \multirow{2}{*}{Execs.} & \multicolumn{2}{c|}{ZTF-SDSS} & \multicolumn{2}{c|}{Gaia-ZTF}  & \multicolumn{2}{c}{AllWISE-SDSS} \\
        \cline{2-7}
        & warm & cold & warm & cold & warm & cold \\
        \hline
        1 & 2885 & 2690 & 3084 & 3115 & 1469 & 1461 \\
        2 & 499 & 1372 & 402 & 1567 & 239 & 745 \\
        4 & 239 & 743 & 209 & 904 & 129 & 436 \\
        8 & 125 & 489 & 123 & 617 & 65 & 268 \\
        12 & 88 & 394 & 77 & 512 & 49 & 198 \\
        16 & 66 & 341 & 60 & 431 & 39 & 162 \\
        20 & 57 & 282 & 51 & 373 & 35 & 140 \\
        24 & 52 & 271 & 45 & 361 & 30 & 129 \\
        28 & 47 & 239 & 39 & 315 & 29 & 122
    \end{tabular}
    \caption{ Averaged raw cross-match performance results (in seconds), when returning all matches, for the last three catalog combinations, depending on the number of executors and whether cold or warm OS cache was used. }
    \label{tblcmraw2}
\end{table}

\begin{table}
    \begin{tabular}{ c | c | c | c | c | c | c }
        \multirow{2}{*}{Execs.} & \multicolumn{2}{c|}{Gaia-AllWISE} & \multicolumn{2}{c|}{Gaia-SDSS}  & \multicolumn{2}{c}{ZTF-AllWISE} \\
        \cline{2-7}
        & warm & cold & warm & cold & warm & cold \\
        \hline
        1 & 723 & 2152 & 642 & 1371 & 3130 & 2991 \\
        2 & 427 & 1184 & 362 & 702 & 363 & 1569 \\
        4 & 197 & 634 & 177 & 362 & 190 & 913 \\
        8 & 134 & 409 & 101 & 217 & 104 & 594 \\
        12 & 74 & 338 & 75 & 196 & 70 & 479 \\
        16 & 60 & 295 & 60 & 181 & 57 & 391 \\
        20 & 57 & 258 & 49 & 160 & 59 & 330 \\
        24 & 47 & 237 & 43 & 156 & 42 & 308 \\
        28 & 42 & 221 & 38 & 147 & 42 & 283
    \end{tabular}
    \caption{ Averaged raw cross-match performance results (in seconds), when returning only the first nearest neighbor, for the first three catalog combinations, depending on the number of executors and whether cold or warm OS cache was used. }
    \label{tblcmrawg1}
\end{table}

\begin{table}
    \begin{tabular}{ c | c | c | c | c | c | c }
        \multirow{2}{*}{Execs.} & \multicolumn{2}{c|}{ZTF-SDSS} & \multicolumn{2}{c|}{Gaia-ZTF}  & \multicolumn{2}{c}{AllWISE-SDSS} \\
        \cline{2-7}
        & warm & cold & warm & cold & warm & cold \\
        \hline
        1 & 2769 & 2617 & 2994 & 2639 & 637 & 1341 \\
        2 & 508 & 1462 & 363 & 1433 & 291 & 730 \\
        4 & 262 & 729 & 607 & 820 & 164 & 417 \\
        8 & 137 & 438 & 111 & 547 & 86 & 256 \\
        12 & 106 & 381 & 77 & 485 & 71 & 198 \\
        16 & 89 & 316 & 64 & 406 & 52 & 164 \\
        20 & 71 & 277 & 56 & 360 & 42 & 142 \\
        24 & 58 & 253 & 47 & 333 & 34 & 130 \\
        28 & 51 & 234 & 42 & 305 & 32 & 116
    \end{tabular}
    \caption{ Averaged raw cross-match performance results (in seconds), when returning only the first nearest neighbor, for the last three catalog combinations, depending on the number of executors and whether cold or warm OS cache was used. }
    \label{tblcmrawg2}
\end{table}

\subsection{Data partitioning performance}
\label{partperf}
We investigated the effects different zone heights and number of buckets have on data preparation. By data preparation we mean sorting and bucketing of the data as was described previously in section \ref{parquet}. This is an operation that needs to be done only once for each catalog (or each new version of a catalog) so that cross-matches can be done online, without additional data movement.

In table \ref{tblrepart} we list the time needed for partitioning each catalog (in minutes) and the size of the partitioned (compressed) Parquet files on disk (in GB), depending on the number of zones used, while using the fixed number of buckets (500, which is the default). We used 28 Spark executors for the tests. The middle columns show the data for the number of zones AXS uses by default (10800 zones, which corresponds to the zone height of one arc-minute). The other two columns show the results for twice as many and half as many zones. As can be seen from the results, the partitioning times depend roughly on the total size of the data but have large fluctuations, so more tests would be needed to make more accurate measurements. However, these numbers are intended to be only informational as users are not expected to need to do this on their own, or at least not often.

Compressed size of partitioned catalogs increases with the number of zones because of increased data duplication, as was explained in section \ref{zoneheightbuckets}. 

\begin{table}
    \begin{tabular}{ c | c | c | c | c | c | c }
        \multirow{2}{*}{Catalog} & \multicolumn{2}{c|}{5400 z.} & \multicolumn{2}{c|}{10800 z.}  & \multicolumn{2}{c}{21600 z.} \\
        \cline{2-7}
        & size & time & size & time & size & time \\
        \hline
        SDSS & 66 & 12 min & 71 & 12 min & 82 & 12 min \\
        Gaia & 430 & 89 min & 464 & 86 min & 532 & 150 min \\
        Allwise & 352 & 120 min & 384 & 119 min & 444 & 133 min \\
        ZTF & 1124 & 547 min & 1169 & 545 min & 1334 & 523 min \\
    \end{tabular}
    \caption{ Data partitioning: size of the partitioned catalogs (in GB) and time needed to partition the data (in minutes) depending on the number of zones used. All tests shown here used 500 buckets for partitioning data. }
    \label{tblrepart}
\end{table}

Table \ref{tblrepartbuck} shows the same tests, but this time depending on the number of buckets used, while using the fixed number of zones (the default of 10800). The middle columns are the same as in table \ref{tblrepart} (they correspond to the same number of buckets and zones). Size of the catalogs is the same regardless of number of buckets used (Parquet compression and size of indexes don't depend on the number of files that much). However, partitioning time obviously decreases with more buckets used. We believe this is because data shuffling is more efficient with smaller files.

\begin{table}
    \begin{tabular}{ c | c | c | c | c | c | c }
        \multirow{2}{*}{Catalog} & \multicolumn{2}{c|}{250 b.} & \multicolumn{2}{c|}{500 b.}  & \multicolumn{2}{c}{750 b.} \\
        \cline{2-7}
        & size & time & size & time & size & time \\
        \hline
        SDSS & 71 & 12 min & 71 & 12 min & 72 & 10 min \\
        Gaia & 464 & 88 min & 464 & 86 min & 464 & 86 min \\
        Allwise & 384 & 125 min & 384 & 119 min & 384 & 116 min \\
        ZTF & 1169 & 557 min & 1169 & 545 min & 1169 & 514 min \\
    \end{tabular}
    \caption{ Data partitioning: size of the partitioned catalogs (in GB) and time needed to partition the data (in minutes) depending on the number of buckets used. All tests shown here used 10800 zones for partitioning data. }
    \label{tblrepartbuck}
\end{table}

\subsection{Cross-matching performance depending on the number of zones and buckets}
We also investigated the effect different number of zones and different number of buckets have on cross-matching performance. Table \ref{tblcrosszones} shows cross-matching performance results when using different numbers of zones while keeping number of buckets fixed at the default value of 500. Table \ref{tblcrossbuckets} shows the same when using different numbers of buckets while keeping number of zones fixed at the default value of 10800. Values in the middle columns in both tables are the same as those in tables \ref{tblcmraw1} and \ref{tblcmraw2} (because those tests used the default values for both the number of zones and the number of buckets). All tests in this section were done using 28 executors and with the queries returning all matching results.

The results show that increasing zone height improves cross-matching performance and that number of buckets doesn't influence cross-matching times much.

\begin{table}
    \begin{tabular}{ c | c | c | c | c | c | c }
        \multirow{2}{*}{Catalogs} & \multicolumn{2}{c|}{5400 z.} & \multicolumn{2}{c|}{10800 z.}  & \multicolumn{2}{c}{21600 z.} \\
        \cline{2-7}
        & warm & cold & warm & cold & warm & cold \\
        \hline
        G - A & 32 & 207 & 31 & 226 & 36 & 240  \\
        G - S & 33 & 128  & 37 & 148 & 36 & 151  \\
        Z - A & 47 & 260  & 38 & 296 & 39 & 283  \\
        Z - S & 48 & 209  & 47 & 239 & 49 & 227  \\
        G - Z & 37 & 271  & 39 & 315 & 44 & 326  \\
        A - S & 27 & 114 & 29 & 122 & 29 & 130  \\
    \end{tabular}
    \caption{ Cross-matching duration (in seconds) depending on the number of zones and whether cold or warm OS cache was used while keeping number of buckets fixed at 500 (the default), for each catalog combination (denoted by their first letters), using 28 executors and returning all results. }
    \label{tblcrosszones}
\end{table}

\begin{table}
    \begin{tabular}{ c | c | c | c | c | c | c }
        \multirow{2}{*}{Catalogs} & \multicolumn{2}{c|}{250 b.} & \multicolumn{2}{c|}{500 b.}  & \multicolumn{2}{c}{750 b.} \\
        \cline{2-7}
        & warm & cold & warm & cold & warm & cold \\
        \hline
        G - A & 32 & 211 & 31 & 226 & 32 & 2314 \\
        G - S & 33 & 146 & 37 & 148 & 37 & 159  \\
        Z - A & 37 & 292 & 38 & 296 & 36 & 289  \\
        Z - S & 47 & 237 & 47 & 239 & 47 & 234  \\
        G - Z & 39 & 323 & 39 & 315 & 40 & 313  \\
        A - S & 28 & 119 & 29 & 122 & 28 & 132  \\
    \end{tabular}
    \caption{ Cross-matching duration (in seconds) depending on the number of buckets and whether cold or warm OS cache was used, while keeping number of zones fixed at 10800 (the default), for each catalog combination (denoted by their first letters), using 28 executors and returning all results. }
    \label{tblcrossbuckets}
\end{table}

\section{Demonstrating a Complete Use Case: Period Estimation for ZTF Light Curves}
\label{sec:gatspy}

Integrating AXS and Spark with astronomical applications and tools will
enable an ecosystem that can query, cross-match, and analyze large
astronomical datasets. We use the example of the analysis of the time
series data described in Section~\ref{sec:org34773cc} to illustrate how existing
tools can be integrated with AXS. For this case we use Gatspy \citep[General
tools for Astronomical Time Series in Python;][]{jake_vanderplas_2015_14833}, a suite of
efficient algorithms for estimating periods from astronomical time-series data. Gatspy is written in Python and contains fast
implementations of traditional Lomb-Scargle periodograms \citep{1976Ap&SS..39..447L, 1982ApJ...263..835S}
and extensions of the Lomb-Scargle algorithm for the case of
multi-period estimation, and multi-band observations. The
implementations of these algorithms scale as O(N) for small numbers of
points within a light curve or times series (i.e. $<10^4$ points) to
O(N$^2$) for large times series. A 10$^3$ point light curve requires
approximately 10$^{-2}$ seconds to estimate the best fitting period.

Python functions are accessed through Spark using user defined
functions (UDFs). There are two basic mechanisms for constructing
UDFs. In each case, a decorator is used to annotate the Python
function (specifying the return type of the output of the
function). The earliest implementation of UDFs serialized and
distributed the data from a Spark or AXS \texttt{DataFrame} to a function
row-by-row. More recently the \texttt{pandas\_udf} has been implemented within
Spark 2.4 which enables vectorized operations. Data partitions are
converted into an Arrow format as columns, serialized, passed to the
Python function, and operated on as a \texttt{Pandas.Series}. Returned data are
converted to an Arrow format and passed back to the Spark driver over a
socket. Because of the vectorized operations and improved
serialization, the \texttt{pandas\_udf} is much more efficient than earlier UDFs \citep{pandas_udf} and it is this
type of UDF that we will consider here. Still, users can expect a degraded performance of UDFs with respect to Spark 
native functions because UDFs are black boxes for Spark Catalyst optimizer. Python UDFs
can add even more performance degradations because of data serialization and deserialization
between different data representations.

For the \texttt{pandas\_udf} there are two \texttt{PandasUDFTypes}, \texttt{SCALAR} and
\texttt{GROUPED\_MAP}. For \texttt{SCALAR} functions the data are passed to and returned
from the Python function as \texttt{Pandas.Series} (with the output being a
\texttt{Pandas. Series} of the same length as the input).  A more flexible
approach is provided by the \texttt{GROUPED\_MAP} type. For this case the spark \texttt{DataFrame}
is subdivided into subsets using a \texttt{groupby} function, serialized and
passed to the Python function using the Arrow format, and the output
returned to the driver as a Spark \texttt{DataFrame}. The flexibility of the
\texttt{GROUPED\_MAP} comes from the fact that arbitrarily complex data
structures can be returned as a \texttt{DataFrame} (e.g. a single period, a
vector of periods and scores for the period, or vector arrays for the
periodogram and associated frequencies). A limitation on the current
\texttt{GROUPED\_MAP} implementation is that the \texttt{apply} function cannot pass arguments to the
Python function.

For our example application we adopt the \texttt{GROUPED \_MAP} data type. The structure of
the \texttt{pandas\_udf} is shown below for a Gatspy function that returns two
arrays, a periodogram and the associated frequencies. The format for
the returned data is described by the schema.

\begin{lstlisting}[language=Python,caption={Applying Pandas functions to partial results},captionpos=b]
schema = StructType(
    [StructField('Frequency',
         ArrayType(DoubleType()),False), 
     StructField('Periodogram',
         ArrayType(DoubleType()),False)])
         
@pandas_udf(schema, PandasUDFType.GROUPED_MAP)
def LombScargle_periodogram(data):
    model = periodic.LombScargle(
        fit_period=True, Nterms=n_term)
    model.fit(data['mjd'], 
        data['psfmag'], data['psfmagerr'])
    periodogram, frequency = 
        model.periodogram_auto()
    return pd.DataFrame(
        dict(Frequency=frequency, 
            Periodogram=periodogram))

results = df.groupby('matchid').
    apply(LombScargle_periodogram)
\end{lstlisting}

\section{Discussion}

\subsection{Similar Systems}
\label{sec:others}
The closest to our approach is a system called ASTROIDE, by \citet{astroide_brahem_2018}. ASTROIDE is also based on Apache Spark and offers an API for cross-matching and processing astronomical data. The main difference compared to AXS is the data partitioning scheme used: AXS partitions the data using a "distributed zones" partitioning scheme, while ASTROIDE uses HEALPix partitioning in a way that closely resembles our prior Large Survey Database product (Section~\ref{sec:prelim_lsd}). Comparing cross-matching performance results between ASTROIDE and AXS is not easy because of the differences in benchmarking environments (ASTROIDE authors used 80 CPU cores spread over 6 nodes, compared to a maximum of 28 cores on a single machine for AXS) and different data sizes (the most relevant ASTROIDE tests were done cross-matching 1.1 billion with 2.5 million objects; the smallest dataset in our tests had $\sim$0.7 billion objects). ASTROIDE shows the best results when cross-matching was performed in advance and saved separately (with "materialized partitions"); we prefer online cross-matching to avoid the $O(N^2)$ problem with given an increased number of datasets available. These differences aside, we find that AXS outperforms ASTROIDE in cross-matching performance (even with materialized partitions): for cross-matching 1.2 billion with 2.5 million objects on 80 cores ASTROIDE needs about 800 seconds. That said, performance was not a main driver in our partitioning approach; AXS has been intentionally designed as a minimal extension to Spark's existing support for SQL and bucketing, rather than a more encompassing rework needed by ASTROIDE. This makes it easier to achieve our goal of upstreaming all Scala-level changes, and eventually having to maintain only the Python API library \footnote{In case our changes do not get merged into the main Spark distribution, maintaining this patch is still a small task relative to building a completely new data management tool (such as QServ).}.

More broadly, the problem of data storage and indexing has been solved within large surveys in various ways. For example, QServ database \citep{qserv_wang_2011} is being built for the LSST project to enable the scientific community to query the LSST data. A unique aspect of QServ is that it implements {\em shared scans}, a technique which allows simultaneous execution of whole-database queries by large numbers of users. QServ doesn't yet implement capabilities for executing more complex data analytics workflows, or image processing functions, something that we've found useful in ZTF work.

The Gaia survey data release pipeline implements a cross-match function described in~\cite{Gaia-cross-match_marrese-at-al_2017}. Their algorithm is based on a sweep line technique which requires the data to be sorted by declination. The data then only needs to be read once. They differentiate between good and bad match candidates and have separate resulting tables for each. They also utilize position errors in determining the best match. The algorithm was implemented in MariaDB, with custom performance optimizations. The authors report cross-match time between Gaia DR1 data set (1.1 billion objects) and SDSS DR9 (470 million objects) of 56 minutes. 

There has been a number of other work focused at providing efficient solutions for cross-matching and handling large astronomical catalogs:
\begin{itemize}
\item \textit{catsHTM} is a recent (2018) tool for ``fast accessing and cross-matching large astronomical catalogs'' ~\citep{catsHTM_soumagnac_2018}. It stores data in HDF5 files and uses Hierarchical Triangular Mesh (HTM) for partitioning and indexing data. They report that cross-matching 2MASS and WISE catalogs takes about 53 minutes (without saving the results).

\item In~\cite{c-m-very_Nieto-santisteban_2018} Nieto-Santisteban et al. describe cross-match implementation in \textit{Open SkyQuery}. It is based on the \textit{zones} algorithm and implemented on Microsoft SQL Server. They report SDSS-2MASS cross-match duration of about 20 minutes when using 8 machines.

\item In~\cite{cross-match-hetero_jia_2015} the authors develop an algorithm for cross-matching catalogs in heterogeneous (CPU-GPU) environments. They index the data using HEALPix indexing method and report the best time of 10 minutes for cross-matching 1.2 billion objects of SDSS data with itself in a multi-node setup with several high-end GPUs.

\item In~\cite{skyQuery_dobos_2012} Dobos et al. implement a probabilistic cross-matching algorithm. They also partition the data based on zones, but each machine contains the full copy of the data. Their cross-match is probabilistic: it is not only based on distances, but on several criteria which all contribute to the final likelihood calculation. They orchestrate multi-node SQL queries using a complex workflow framework, but do not report any performance numbers.

\end{itemize}

\subsection{Summary and Future Directions}
\label{sec:conclusion}

In this paper we presented Astronomy eXtensions for Spark, or AXS, a data management solution capable of operating on distributed systems and supporting complex workflows intermixing SQL statements and Python code. AXS enables scalable and efficient querying, cross-matching, and analysis of astronomical datasets in the $O(10^10)$ row regime. It is built on top of Apache Spark, with minimal extensions to the core Spark code that have been submitted for merging upstream. We described AXS' data partitioning and indexing scheme and the applied epsilon join optimization which enable fast catalog cross-matching and reduce data skew. We also described AXS Python API which exposes AXS cross-matching, light-curve querying functions, histograming, and other functionality. The cross-matching performance testing results show that AXS outperforms other systems, with hardware differences taken into account. The tests done so far have all been performed on a single large machine; work is ongoing to deploy and benchmark AXS in a fully distributed setting.

As we discussed in Section~\ref{sec:org197424c}, the exponential growth of data being generated by large astronomical surveys, and the shifting research patterns towards statistical analyses and examinations of whole datasets present challenges to management of astronomical data in classic RDBMS systems. We argue that systems like Spark and AXS, or perhaps in the future similar systems built on Dask, may serve as a capable, open-source, solution to the impending crisis.

More broadly, the scale of future datasets and the demand for large-scale and complex operations on them poses significant challenges to the typical ``subset-download-analyze'' analysis paradigm common today. Rather than downloading (now large) subsets, there are strong arguments to ``bring the code to the data'' instead, and remotely perform next-to-the-data analysis via {\em science platforms} (e.g., Juric, Dubois-Felsmann \& Ciardi 2016). However, this would place new demands on astronomical archives: the kinds of analyses and the size of the community to be supported would require petascale-level end-user compute resources to be deployed at archive sites. Furthermore, to enable efficient joint analyses / dataset fusion, large datasets of interest would eventually need to be replicated across all major archives, adding to storage requirements. Taking this route is possible, but brings with it all operational issues typically encountered in user-facing HPC deployments (in addition to the broader question of utilization and cost-effectiveness). Is turning astronomical archives into large data-center operators the right way to go?

An alternative may be to consider migrating away from the traditional RDBMS-backends, and towards the management of large datasets using solutions such as AXS, built on scalable, industry-standard, data processing frameworks that map well to the types of science analyses expected for the 2020s. As these solutions are designed to operate in distributed, cloud, environments, so they would enable utilizing the cloud to satisfy the new computational demands. In the best case, the shift would be dovetailed by a move towards physical co-location of datasets on public cloud resources as well (or a hybrid, private-public, solution, that would allow the analysis to begin at archive centers, but then effortlessly spill over into the cloud). Such {\em cloud-native} approach would offer tremendous benefits to researchers: elasticity and cost effectiveness, scalability of processing, shareability of input datasets and results, as well as increased reproducibility.

\acknowledgments

P.~Zecevic, C.T.~Slater, M.~Juric and A.J~Connolly acknowledge support from the University of Washington College of Arts and Sciences, Department of Astronomy, and the DIRAC Institute. University of Washington's DIRAC Institute is supported through generous gifts from the Charles and Lisa Simonyi Fund for Arts and Sciences, and the Washington Research Foundation. 

P.~Zecevic and S.~Loncaric acknowledge support from University of Zagreb Faculty of Electrical Engineering and Computing and the European Regional Development Fund under the grant KK.01.1.1.01.0009 (DATACROSS).

M.~Juric acknowledges the support of the Washington Research Foundation Data Science Term Chair fund, and the UW Provost's Initiative in Data-Intensive Discovery.

A.J~Connolly acknowledges support from NSF awards AST-1715122 and OAC-1739419,

Based on observations obtained with the Samuel Oschin Telescope 48-inch and the 60-inch Telescope at the Palomar Observatory as part of the Zwicky Transient Facility project. ZTF is supported by the National Science Foundation under Grant No. AST-1440341 and a collaboration including Caltech, IPAC, the Weizmann Institute for Science, the Oskar Klein Center at Stockholm University, the University of Maryland, the University of Washington, Deutsches Elektronen-Synchrotron and Humboldt University, Los Alamos National Laboratories, the TANGO Consortium of Taiwan, the University of Wisconsin at Milwaukee, and Lawrence Berkeley National Laboratories. Operations are conducted by COO, IPAC, and UW.

\bibstyle{aasjournals}

\bibliography{axsrefs.bib}

\end{document}